\documentclass[twocolumn,amsmath,amssymb,aps,prl,10pt]{revtex4}
\pdfoutput=1

\usepackage{graphicx}
\usepackage{color}
\usepackage{amsmath}
\newcommand{\ket}[1]{\left | \, #1 \right \rangle}
\newcommand{\kets}[1]{| \, #1 \rangle}

\newcommand{\av}[1]{\langle\, #1\,\rangle}
\newcommand{\bracket}[3]{\left\langle\, #1\,|#2|\,#3\,\right\rangle}
\newcommand{\brackets}[3]{\langle\, #1\,|#2|\,#3\,\rangle}

\begin{document}
\title{Entanglement growth in many-body localized systems with long-range interactions}
\author{M. \surname{Pino}}
\affiliation{Sorbonne Universit\'es, UPMC Univ Paris 06, UMR 7589, LPTHE, F-75005, Paris, France}
\affiliation{CNRS, UMR 7589, LPTHE, F-75005, Paris, France.}

\begin{abstract}

We study the time evolution of one-dimensional systems of fermions with long-range interactions in the presence of strong disorder. Exact diagonalization of small systems supports many-body localization for weak Coulomb and dipolar interactions at high energy. The dephasing time of far apart degrees of freedom is analyzed. The result indicates that von Neumann entropy of one halve of the system evolves as a power law on time for a disentangled initial state. The numerical simulations performed in the Coulomb and dipolar cases are consistent with this prediction.

\end{abstract}
\maketitle

Interactions often affect the behaviour of a system in interesting ways.
Basko \textit{et al.} shown that an isolated system of particles in a one-dimensional lattice with disorder 
may undergo a metal-insulator transition\ \cite{BaAl06}. 
This only  happens in the presence of interactions, because all the single-particle  states are localized for any non-zero disorder\ \cite{AbAn70}.
Numerical evidences of this many-body transition were reported for spins models\ \cite{CuFe12}.
Furthermore, the localized phase could persist at infinite temperature for strong enough disorder\ \cite{PaHu10}. 
Many-body localized states at high temperature exhibit properties which are rather different than its 
non-interacting counterparts.

A closed system of particles which are localized cannot thermalize. Indeed, any excess of matter, or energy, does not diffuse under unitary time evolution.
This lack of thermalization is expected to have consequences in the time evolution of the von Neumann entropy of one halve of the system. 
Indeed, the infinite time entropy saturates with the size of the system in the absence of interactions\ \cite{BaPo12}.
Intuitively, localization prevents the system to explore a significant part of the available states in the microcanonical ensemble. 

Many-body localized system with nearest-neighbors interactions behaves in a different way.
The evolution of one halve entropy from an initial disentangled state grows logarithmically 
on time up to a value proportional to the system size\ \cite{ChMo06,ZniPr08,BaPo12,VoAl13,SePa13}.
It has been argued that the proportionality constant is small to allow thermalization\ \cite{BaPo12,RiDu08}. 
One may also think that a logarithmic growth is rather slow for an usual thermal state.
Here, we show that long-range interactions produce a faster, power law, evolution of the entropy in a localized system.
Our numerical results also indicates that the entropy saturates with time to a value which is the same for local and long-range many-body effects.

A crucial point for our work is whether many-body localization exists at all in the presence of long-range interactions. 
Recently,  Yao \textit{et al.} analyzed the number of resonances of spin systems\ \cite{YaLa13}.  
Regarding nearest-neighbors hopping, they conclude the existence of a many-body localized phase for Coulomb and dipolar interactions in one dimension.
This is supported by their numerical data in the dipolar case.
We will present results which also indicates a many-body localized phase for both types of interactions at high energy.

The evolution of entropy has been analyzed for models with Ising- and XXZ-like long-range interactions, but without disorder\ \cite{DuHa05, EiW013}. 
In those cases, entanglement growth is usually due to pseudo-particles carrying information between far apart regions\ \cite{HaTa13}.
We notice that, even in the case of delocalized excitations, entropy spread could be faster than pseudo-particle propagation as reported by Huse\ \cite{KiHu13}.
Pseudo-particles cannot be the responsible of entropy growth when localization takes place. 
Thus, it is remarkable that we find a fast spread in many-body localized system.

The experimental observation of many-body localization would provide with an experimental ground for all the analytical and numerical results obtained in the last years\ \cite{NaGo14}.
A recent proposal, which uses the entropy evolution, 
is argued that can provide experimental evidences of the phenomenon under consideration\ \cite{SeGo14}.
In such an implementation, it is important to take into account that weak long-range interactions lead to a quicker dephasing, 
specially when the localized degrees of freedom are electrons. In this case, Coulomb interactions are important due to the lack of screening.
The results reported here can be tested experimentally using similar methods to the ones appearing in Ref.\ \cite{DaPi13,JuLa14}, where the evolution of entropy and correlations is directly measured. 

\subsection{Many-body localization with long-range interactions}

We consider fermions in a one-dimensional lattice with $L$ sites. The Hamiltonian is $H=H_0+H_I$, where:
\begin{align}
 H_0&=-\sum_{\av{i,j}} c_{i}^{\dagger}c_{j}+c_{j}^{\dagger}c_{i}
+\sum_{i} \phi_{i}\ n_{i},\nonumber\\
H_{\rm I}&=V\sum_{i<j}\frac{n_i\ n_j}{|r_i-r_j|^p}. \nonumber
\end{align}
The first summation in $H_0$ runs over nearest-neighbors sites. The strength of the interactions is $V$ and the $\phi's$ are on-site potentials, which are randomly distributed in $[-W,W]$.
The  $c_{i}^{\dagger}$, $c_{i}$ and $n_i=c^\dagger_{i}c_{i}$ are the usual fermionic creation, annihilation and number operators
in the site $i$ of the lattice, respectively. The decay of the interactions is controlled by $p$.
In the numerical simulations, we use disorder $W=8$, open boundaries conditions and half occupation.
The lattice constant is $1$ and units in which $\hbar=1$ are taken.
The Hamiltonian used here maps to a spin $1/2$ system with long-range coupling via Jordan-Wigner.

The existence of many-body localized eigenstates at high energy is first explored.
The eigenstate with energy closer to zero, $\kets{\psi_h}$, are computed via exact diagonalization for system of sizes up to $L=14$. 
Several realizations of the disorder are averaged. 
In Fig.\ (\ref{Fig1}), the logarithmic of the two body 
correlations, $\ln(\brackets{\psi_h}{c_L c_1^\dagger}{\psi_h})$,  between the first and last site are shown as a function of the system size. The disorder is $W=8$. The circles correspond to $V=0$.  
The other symbols represent weak interacting systems, $V=0.1$, for several types of interactions:  $p=\infty$ (nearest-neighbors), $p=1$ and $p=3$, as indicated in the legend.
The results are pretty much the same indicating that interactions do not delocalize the fermions. Indeed, the localization length, which can be obtained from the decay of the correlations, is the same 
for all the cases $\xi\sim 1$.
As in Ref.\ \cite{YaLa13}, our numerical simulations support the existence of a many-body localized phase for long-range interactions for $p=1,3$.

\subsection{Entanglement growth}

\begin{figure}[t!]
  \includegraphics[height=.24\textheight]{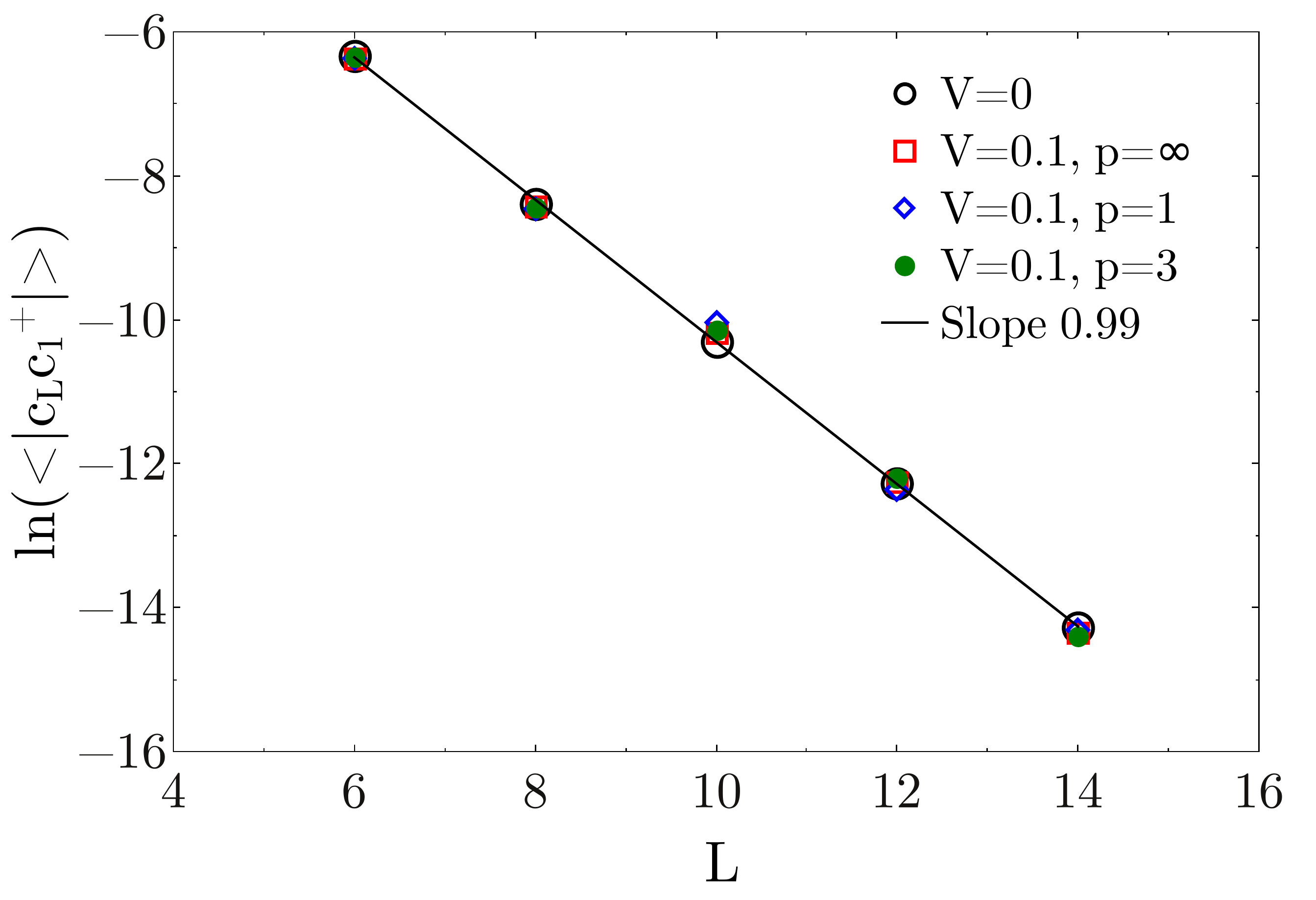}
  \caption{ The logarithm of two points correlations for the eigenstate with energy closer to zero, $\ln(\brackets{\psi_h}{c_L c_1^\dagger}{\psi_h})$, appears as a function of the size of the system, $L$.
The disorder is $W=8$.
The case for no interactions are represented as $\circ$. The many-body cases correspond each to: $\square$ for nearest-neighbors, $\diamond$ Coulomb and $\bullet$ for dipolar interactions. 
The strength of interactions is $V=0.1$. The solid line is a lineal fit of the data for $V=0$. The slope is approximately $1$. The data are averaged over 1000 samples.} \label{Fig1}
\end{figure}

We study the regime in which all the eigenstates of the non-interacting system are strongly localized, with localization length similar to the lattice spacing, $\xi\sim 1$.
Weak interactions, $V\ll 1$, are taking into account using first order perturbation theory. 
We saw in Fig.\ (\ref{Fig1}) that the localization length of the high energy eigenstates are the same for the non-interacting and interacting cases, 
regardless of the shape of many-body effects. We have obtained the same results for other energies. 
Thus, the structure of the non-interacting eigenstates  does not seem to be affected by weak interactions in the strongly localized regime.

The eigenstates  for $V\ll 1$  are approximated by the ones of the non-interacting problem. Those are Slater determinants
$\kets{\varphi_{\tilde{\alpha}}}=\Pi_{i=1}^{m} \kets{\phi_{\alpha_i}} $, where $\kets{\phi_{\alpha_i}}$ are single-particle eigenstates and $m$ the number of particles.  
The tilde means a multiple index $\tilde{\alpha}=\left\lbrace\alpha_1,\dots, \alpha_m\right\rbrace$.
The first order correction to the eigenenergies are $E_{\tilde{\alpha}}=E_{\tilde\alpha}^{(0)}+\sum_{i<j}\delta E_{\alpha_{i}\alpha_{j}}$,
where $E_{\tilde\alpha}^{(0)}$ is the non-interacting energy and $\delta E_{\alpha\beta}=\bracket{\phi_{\alpha}\phi_{\beta}}{H_{\rm I}}{\phi_{\alpha}\phi_{\beta}}$.
We notice that $\kets{\phi_{\alpha}\phi_{\beta}}= a^{\dagger}_{\alpha}a^{\dagger}_{\beta}\kets{0}$, where the symbols $a^{\dagger}_{\alpha}$ 
are the creation operators in the base of single-particle eigenstates.

We analyze the time evolution of two far away particle in an initial disentangled state: $\kets{\Psi}=\kets{\psi_1\psi_2}$. 
We suppose that the localization length of this state is similar to the one of the eigenstates of the system, $\xi$. 
Time evolution from the initial state is $\kets{\Psi}=\sum\lambda^\alpha_1\lambda^\beta_2 e^{-iE_{\alpha \beta}}\kets{\phi_\alpha\phi_\beta}$, 
where $\kets{\psi_i}=\lambda^{\alpha}_i\kets{\phi_\alpha}$.
We denote the expected position of each of the particles as $x_i=\bracket{\psi_i}{\hat{r}}{\psi_i}$, where $\hat{r}$ is the position operator. 
The particles are localized far apart $R=|x_2-x_1|\gg\xi$.
Then, $\lambda^\alpha_i\neq 0$ if the corresponding eigenstate $\kets{\phi_\alpha}$ is localized near $x_i$. 
We compute the first order correction to the energies of $\kets{\phi_\alpha\phi_\beta}$, when $\kets{\phi_\alpha}$ is localized near $x_1$ and $\kets{\phi_\beta}$ around $x_2$. 
Exponentially small contribution $\exp(-R/\xi)$ are neglected.
Using a multipole expansion:
\begin{align}
 \delta E_{\alpha\beta}\approx V\left(\frac{1}{R^p}-p\frac{\mu_{\alpha}-\mu_{\beta}}{R^{p+1}}+\frac{p(p+1)\mu_{\alpha}\mu_{\beta}}{R^{p+2}}\right),
\end{align}
where $\mu_{\alpha}=\bracket{\phi_{\alpha}}{\hat{r}}{\phi_{\alpha}}$. The last term is analogous to a dipole-dipole interaction and it is the unique one
that entangles the state of the two particles during time evolution.
The dephasing time  depends on the distance as $R^{p+2}$.
That is, the two particle become effectively entangled only after a time $t\sim R^{p+2}$. 
The maximum von Neumann entropy of each of the particles will also occur at that time. 

Our analysis is similar to the one performed for local interactions in Ref.\ \cite{SePa13}, in which 
 the corrections to the relevant energies are $ \delta E_{\alpha\beta}\approx Vb_{\alpha\beta}e^{-R/\xi}$ with $b_{\alpha\beta}$ real numbers.
In this case, the dephasing time increases exponentially with the distance between particles. 
We see that long-range interactions produce a much quicker dephasing and then, faster entropy growth.

We investigate time evolution of a particle in the left halve of the system, in a state $\ket{\psi_1}$, which interacts with many other far apart localized particles 
at the right halve, which are in states $\ket{\psi_j}$ for $j>1$.
That is, we consider $m$ fermions in an initial state $\kets{\Psi}=\Pi_{i=1}^{m}\kets{\psi_i}$. The positions $x_i$ are defined as before. We 
assume the particle at left is far from the ones on the right halve $R_j=|x_1-x_j|\gg\xi$,  for $j>1$. 
The time evolution is $\kets{\Psi}=\lambda^\alpha_1\lambda^{\tilde{\beta}} e^{-iE_{\alpha \tilde\beta}}\kets{\phi_\alpha\varphi^{m-1}_{\tilde\beta}}$, 
where $\lambda^{\tilde{\beta}}=\lambda^{\beta_2}_2\dots\lambda^{\beta_m}_m$.
The energy of $\kets{\phi_\alpha\varphi^{m-1}_{\tilde\beta}}$ is up to first order $E_{\alpha,\tilde\beta}=E_\alpha^{(0)}+E_{\tilde\beta}+\sum_{j>1} \delta E_{\alpha\beta_j}$.
Using similar arguments as for the evolution of two particles: 
\begin{align}
\sum_{j>1} \delta E_{\alpha\beta_j}\approx V\sum_{j>1}\left(\frac{1}{R_j^p}-p\frac{\mu_{\alpha}-\mu_{\beta_j}}{R_j^{p+1}}+\frac{p(p+1)\mu_{\alpha}\mu_{\beta_j}}{R_j^{p+2}}\right),
\end{align}
Again, only the analogous to dipole-dipole interaction entangles the particle at left with the rest of the system, but now this particle interacts with many other. 
A estimation of the dephasing time can be obtained by taking an uniform distribution of particles on the right halve. 
Then, the summation of the dipole-dipole interactions can be performed as an integral. 
For a large system, we obtain a rough estimation of the dephasing time as
$t_{dep}\sim R_{12}^{p+1}$, where $R_{12}$ is the distance between particle at left and the first particle at right.

Now, we can understand the general evolution starting from a disentangled state $\ket{\Psi}=\ket{1010\dots10}$. 
As the strongly localized regime is considered, this state can be approximated as a Slater determinant in which each electron is 
in a superposition of a few nearby localized single-particle eigenstate. 
After a time $t$, each particle closer to the right halve than a distance $x(t)$, which is of the order of $x(t)\sim t^{\frac{1}{p+1}}$, is effectively entangled. 
Assuming that the entropy of the left halve of the system is proportional to the number of particles entangled with the right halve, we deduce:
\begin{align}\label{eq:S}
S(t)\sim t^{\frac{1}{p+1}}.
\end{align}
This formula has been obtained by analyzing the most relevant time scales in which dephasing occurs. 
Arguments based on the density matrix of one halve of the system can be also employed as in Ref.\ \cite{SePa13}.
In the following, we analyze the validity of this law using numerical simulations.

\subsection{Numerical results}

\begin{figure}[t!]
\hspace{-0.5cm}  \includegraphics[height=.24\textheight]{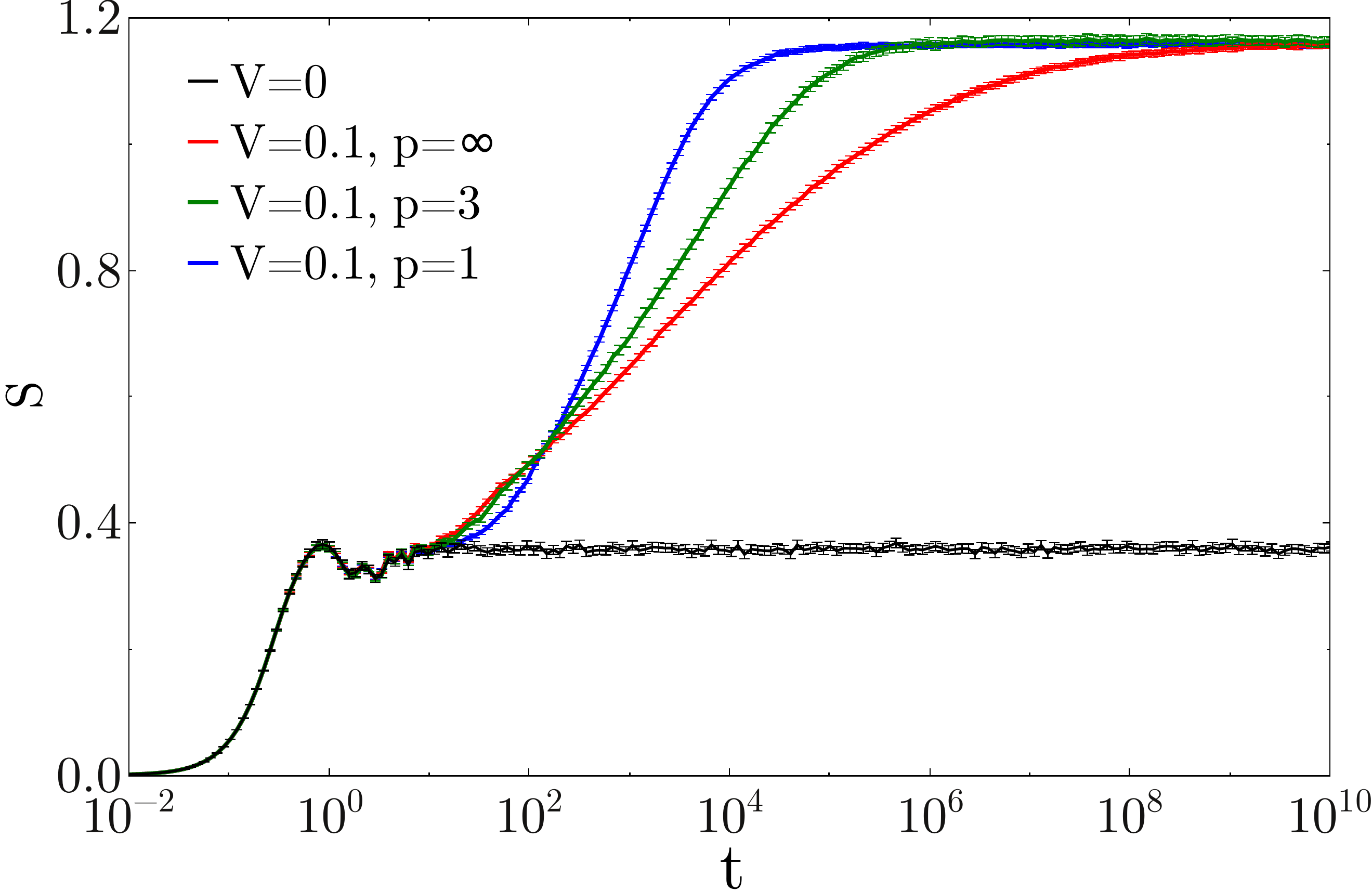}
  \caption{ Time evolution of Von Neumann entropy, $S$ from an initial state $\ket{\Psi}=\ket{1010\dots10}$ in a system of size $L=14$. 
The disorder is $W=8.0$. The black line corresponds to the non-interacting case, red line represents the data for nearest-neighbors interactions ($p=\infty$), 
green line is used for the case of dipolar interactions and the blue one is used for Coulomb ones. Interactions are weak, $V=0.1$. Average over, at least 1000 samples have been performed.
} \label{Fig2}
\end{figure}

Time evolution from an initial disentangled product state $\ket{\Psi}=\ket{1010\dots10}$ is analyzed.
Exact diagonalization of the Hamiltonian, with disorder $W=8$, is performed at half filling. 
Time evolution is then computed using the eigenvectors and eigenvalues.

In Fig.\ (\ref{Fig2}), the von Neumann entropy, $S$, is shown as a function of time, $t$, for systems of size $L=14$. 
Each of the lines corresponds to non-interacting fermions, nearest-neighbors, dipolar and Coulomb interactions, as specified in the legend. 
The interaction strength is $V=0.1$. The evolution is the same for all the cases shown at small $t$. 
Indeed, diffusive spreading of fermions up to the localization length takes place at those early times. 
After, $S$ grows in a different way for each case. 
Entropy saturates in non-interacting systems. A $\log(t)$ growth can be seen in the presence of nearest-neighbors interactions, $p=\infty$.
Coulomb ones are characterized by a faster increase of $S$ than a logarithmic. The case of dipolar interactions
is more difficult to analyze, though S grows faster than for nearest-neighbors ones.
All of this is consistent with our previous analysis.

In Fig.\ (\ref{Fig2}), entropy saturates to the same value for $p=1,3$ and nearest-neighbors interactions. We have checked that this also occurs for any of the accessible sizes. 
Thus, saturation of entropy depends linearly on the subsystem size, as it has been found for local interactions.
It should be noticed that entropy at infinite time depends on the initial state\ \cite{SePa13}.
We have also seen that particle number fluctuations behaves in a similar way as in 
the case of local many-body effects. That is, it saturates with time to a value which varies very little with system size, which implaies insulating behaviour. 

\begin{figure}[t!]
  \includegraphics[height=.24\textheight]{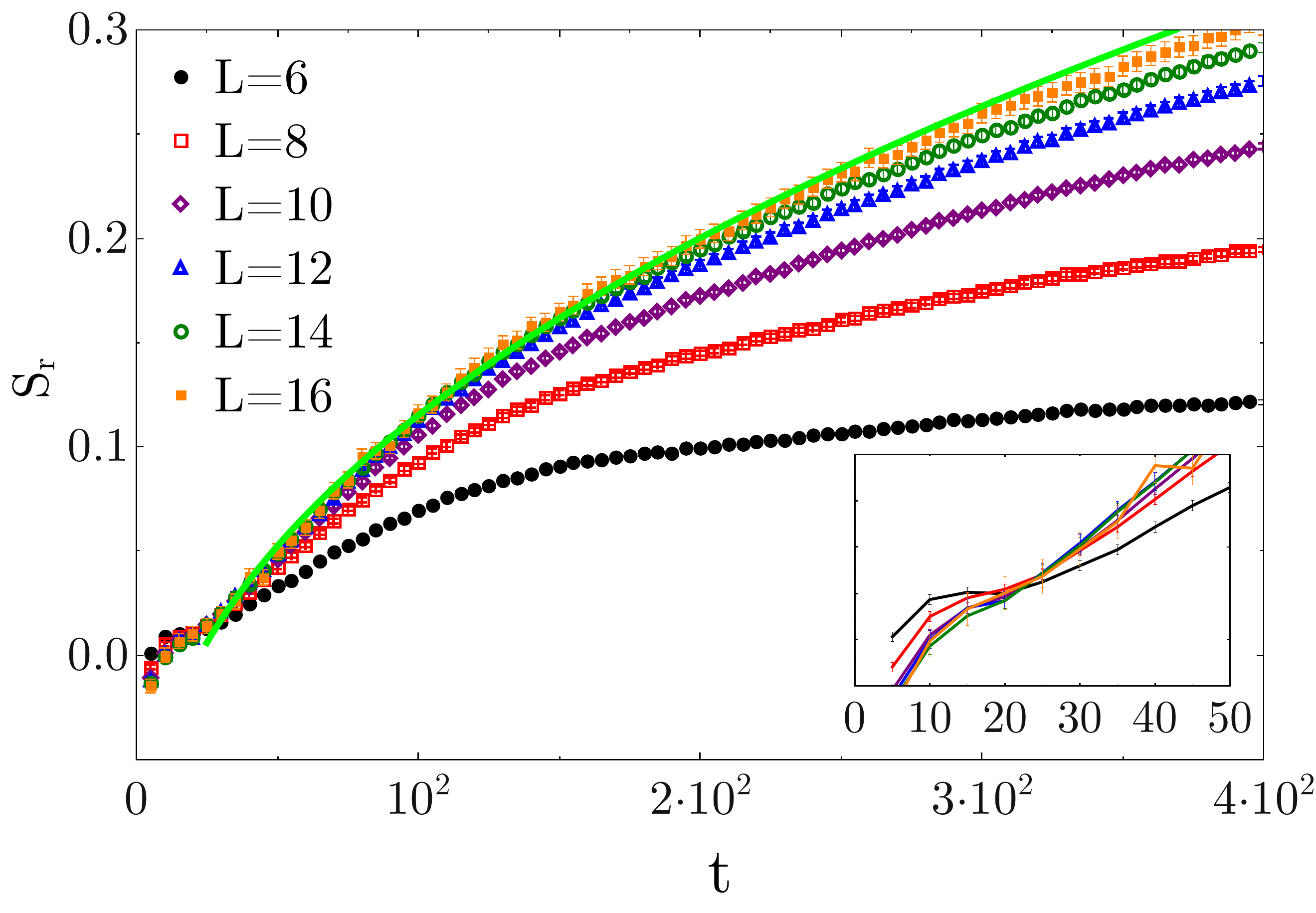}
  \caption{Rescaled von Neumann entropy $S_r(t)$, which is the difference between the interacting entropy and saturation value of the non-interacting case,  as a function of time. 
The initial state  is: $\ket{\Psi}=\ket{1010\dots10}$. The disorder is $W=8.0$
 Each data corresponds to sizes from $L=6$ (bottom), to $L=16$ (top). Average over, at least, 4000 samples have been performed.
The green solid line is the result of fitting the $L=16$ curve to a law $at^q+b$, which results in $q\sim 0.44\pm0.03$. 
This fit has been performed in the time interval $[25,240]$. Inset: zoom of the same data from $t=0$ to $t=50$.
} 
\label{Fig3}
\end{figure}

It is difficult to deduce the functional form of $S(t)$ from the data in Fig.\ (\ref{Fig2}). Finite size effects are expected to become more severe in the case of long-range interactions. 
We perform a finite size analysis for weak Coulomb interactions, $V=0.1$. 
In small systems, the saturation  entropy of non-interacting fermions depends weakly on the system size, which makes finite size analysis even  more complicated. 
For this reason, the entropy $S(t)$ has been rescaled for each size as $S_r(t)=S(t)-S_0(\infty)$, where $S_0(\infty)$ is 
the saturation value of entropy without interactions. Average over, at least $4000$ samples have been taken.
Rescaled entropy, $S_r$, as a function of time appears in Fig.\ (\ref{Fig3}) for sizes $L=6,8,10,12,14$ and $16$. 

The re-scaled entropy of small system is underestimated at large times due to finite size corrections.
The data for sizes $L=14$ and $L=16$ are not compatible for the first time at $t=240$.
Finite size effects are relevant after that time for the curve $L=16$. 
At early times, the evolution of entropy, $S$, is not due to many-body effects. 
It is because of diffusion at small distances and does not depend on the size of the system. 
However, $S_0(\infty)$ slightly increases with system size. Thus, the re-scaled entropy tends to be larger for small systems at small $t$. 
A crossing point appears near $t=20$ (inset of Fig.\ (\ref{Fig3})), which indicates that the evolution of $S_r(t)$ is due to many-body effects after that time.
We perform a fit to a polynomial law $at^q+b$ in the interval $t=25$, to $t=240$ for the curve $L=16$, which gives $q=0.44\pm0.03$
This result is consistent with our earlier predictions given the limitation in the  sizes that we can handle. 

\section{Conclusions}

Entropy growth in a many-body localized system with long-range interactions is faster than in the case of local ones.
A power law $t^{\frac{1}{p+1}}$, where $p$ controls the spatial decay of the many-body effects, is found using arguments based on dephasing of far apart degrees of freedom.
Our simulations for system of sizes up to $L=16$ fit well this prediction for Coulomb interactions and 
they are also consistent for dipolar ones. The entropy saturates at large time to a value which turns out to be independent of the type of interactions. 
That value is proportional to the system size but, as in the case of local many-body effects, it is small to allow thermalization\ \cite{BaPo12}.
The fast entanglement growth reported here should be taken into account in experiments dealing with many-body localization in which interactions are long-ranged. 
For example, in the case of dipolar molecules in optical lattices.

\end{document}